\documentclass[aps,
superscriptaddress,amsmath,amssymb]{revtex4}
\usepackage{graphicx}
\usepackage[dvips]{color} 
\begin{document}

\title{Separability criteria with angular and Hilbert space averages}

\author{Kazuo Fujikawa}
\affiliation{Quantum Hadron Physics Laboratory, RIKEN Nishina Center, Wako 351-0198, Japan}
\affiliation{Institute of Advanced Studies, Nanyang Technological University, Singapore 639798, Singapore}
\author{C.H. Oh}
\affiliation{Institute of Advanced Studies, Nanyang Technological University, Singapore 639798, Singapore}
\affiliation{Centre for Quantum Technologies, National University of Singapore, Singapore 117543, Singapore}
\author{Koichiro Umetsu}
\affiliation{Laboratory of Physics, College of Science and Technology, Nihon University, Funabashi 274-8501, Japan}
\author{Sixia Yu}
\affiliation{Centre for Quantum Technologies, National University of Singapore, Singapore 117543, Singapore}

\begin{abstract}
The practically useful criteria of separable states $\rho=\sum_{k}w_{k}\rho_{k}$ in $d=2\times2$ are discussed.
The equality $G({\bf a},{\bf b})= 4[\langle \psi|P({\bf a})\otimes P({\bf b})|\psi\rangle-\langle \psi|P({\bf a})\otimes{\bf 1}|\psi\rangle\langle \psi|{\bf 1}\otimes P({\bf b})|\psi\rangle]=0$ for any two projection operators $P({\bf a})$ and $P({\bf b})$ provides a necessary and sufficient separability criterion in the case of a separable pure state $\rho=|\psi\rangle\langle\psi|$. We propose the separability criteria of mixed states, which are given by ${\rm Tr}\rho\{{\bf a}\cdot {\bf \sigma}\otimes {\bf b}\cdot {\bf \sigma}\}=(1/3)C\cos\varphi$ for two spin $1/2$ systems and $4{\rm Tr}\rho \{P({\bf a})\otimes P({\bf b})\}=1+(1/2)C\cos2\varphi$ for two photon systems, respectively, after taking a geometrical angular average of  ${\bf a}$ and ${\bf b}$ with fixed $\cos\varphi={\bf a}\cdot{\bf b}$. Here $-1\leq C\leq 1$, and the difference in the numerical coefficients $1/2$ and $1/3$ arises from the different rotational properties of the spinor and the transverse photon. If one instead takes an average over the states in the $d=2$ Hilbert space, the criterion for two photon systems is replaced by $4{\rm Tr}\rho \{P({\bf a})\otimes P({\bf b})\}=1+(1/3)C\cos2\varphi$. 
 Those separability criteria are shown to be very efficient using the existing experimental data of  Aspect et al. in 1981 and Sakai et al. in 2006. When the Werner state is applied to two photon systems, it is shown that the Hilbert space average can judge its inseparability but not the geometrical angular average.
\end{abstract}

\maketitle

\large
\section{Introduction}
To characterize the entanglement, the notions of separability and inseparability, which are the characteristic properties of state vectors in quantum mechanics, are commonly used. On the other hand, the notion of local realism based on local non-contextual hidden-variables models is also used to test the properties  such as locality and reduction~\cite{bell, chsh}. Local hidden-variables models are generally different from quantum mechanics, and thus local realism tests the deviation from quantum mechanics also.

In the experimental study of local realism, which is commonly tested by CHSH inequality~\cite{chsh},
it is customary to first confirm the consistency of the measured basic correlation such as the spin correlation $\langle {\bf a}\cdot{\bf \sigma}\otimes {\bf b}\cdot{\bf \sigma}\rangle$  with quantum mechanics and then test the CHSH inequality. If one confirms the consistency
of $\langle {\bf a}\cdot{\bf \sigma}\otimes {\bf b}\cdot{\bf \sigma}\rangle$ for any 
unit vectors ${\bf a}$ and ${\bf b}$ with quantum mechanics, one can naturally apply the criterion of quantum mechanical separability to the correlation. In the present study, we discuss this aspect of separability test in quantum mechanics. It is well known that the Peres criterion of the positivity of partial transposed density matrix gives a necessary and sufficient condition of separability of general density matrix in $d=2\times2$~\cite{peres} which we study in the present paper. However, it is also well-known from the days of Pauli~\cite{weigert} that the reconstruction of
the state vector or density matrix from measured data is in general very involved~\cite{james}. It is thus practically useful to derive simpler criteria which do not require the precise state reconstruction. The purpose of the present paper is to derive such separability criteria.  We assume the most general separable density matrix but we use a limited set of two-point correlations and thus obtain only the necessary condition of separability in general. Nevertheless, we illustrate that our criteria are very useful when applied to the past experimental data of Aspect et al. in 1981~\cite{aspect1} and  Sakai et al. in 2006~\cite{sakai}.

To be specific, we study the general separable quantum mechanical states 
\begin{eqnarray}
\rho=\sum_{k}w_{k}\rho_{k},
\end{eqnarray} 
in $d=2\times2=4$; all the states $\rho_{k}$ are separable pure quantum states~\cite{werner}.
If the state is a separable pure state $\rho=|\psi\rangle\langle\psi|$, separability in (1) is quantified by the equality, which is necessary and sufficient,
\begin{eqnarray}
G({\bf a},{\bf b})
\equiv 4[\langle \psi|P({\bf a})\otimes P({\bf b})|\psi\rangle
-\langle \psi|P({\bf a})\otimes{\bf 1}|\psi\rangle\langle \psi|{\bf 1}\otimes P({\bf b})|\psi\rangle]=0, 
\end{eqnarray}
for arbitrary two projection operators $P({\bf a})$ and $ P({\bf b})$. 

In the case of a general mixed  $\rho$, 
we derive the useful criteria of separable mixed states, namely
\begin{eqnarray}
&&{\rm Tr}\rho\{{\bf a}\cdot {\bf \sigma}\otimes {\bf b}\cdot {\bf \sigma}\}=(1/3)C\cos\varphi,
\end{eqnarray}
for two spin $1/2$ systems, and 
\begin{eqnarray}
4{\rm Tr}\rho \{P({\bf a})\otimes P({\bf b})\}=1+(1/2)C\cos2\varphi, 
\end{eqnarray}
for two photon systems with linear polarization projectors $P({\bf a})$ and $P({\bf b})$, respectively: Here $-1\leq C\leq 1$. The basic new ingredient in our derivation of these criteria (3) and (4) is that we take a geometric {\em angular average} of unit vectors ${\bf a}$ and ${\bf b}$ with fixed $\cos\varphi={\bf a}\cdot{\bf b}$. It is shown that this angular averaging, which is originally motivated by the specific experiment~\cite{sakai}, does not add a new burden to measurements by analyzing the existing experimental data. The difference in the numerical coefficients $1/3$ and $1/2$ in (3) and (4) arises from the difference in  rotational properties; the spinor rotational freedom is 3-dimensional, which agrees with the freedom of the $d=2$ Hilbert space,  while the rotational freedom of the photon is  two-dimensional, which differs from the freedom of the $d=2$ Hilbert space since it is confined in a plane perpendicular to the momentum direction. If one instead takes an average over the states in the $d=2$ Hilbert space, the formula for the photon is replaced by 
\begin{eqnarray}
4{\rm Tr}\rho \{P({\bf a})\otimes P({\bf b})\}=1+(1/3)C\cos2\varphi. 
\end{eqnarray}
This difference between the geometrical angular average and the Hilbert space average is interesting, and it has an interesting implication on the separability
issue of the Werner state~\cite{werner} which accommodates a specific local hidden-variables representation and thus satisfies CHSH inequality. In the case of photon, the Hilbert space average can judge the inseparability of the Werner state but the geometrical angular average cannot.

\section{Criteria of separability}
We discuss the separability criterion on the basis of explicit experimental
data by Aspects et al. in 1981~\cite{aspect1} and Sakai et al. in 2006~\cite{sakai}. In these experiments, the authors emphasize a good agreement of the measured basic correlation such as $\langle P({\bf a})\otimes P({\bf b})\rangle$ or the corresponding quantity for spin operators for any unit vectors ${\bf a}$ and ${\bf b}$ with the predictions of quantum mechanics and then discuss the test of the CHSH inequality. We instead formulate the separability criterion for the two-point function
\begin{eqnarray}
{\rm Tr}\rho \{P({\bf a})\otimes P({\bf b})\},\nonumber
\end{eqnarray}
which inscribes much information about quantum mechanics in $d=2\times2$, and discuss the entanglement from a point of view of separability.
\\

\noindent {\bf Separability criterion of pure states}:\\

We begin with the experimental analysis of separability for a pure state.
\begin{figure}[h]
  \begin{center}
    \includegraphics[width=80mm]{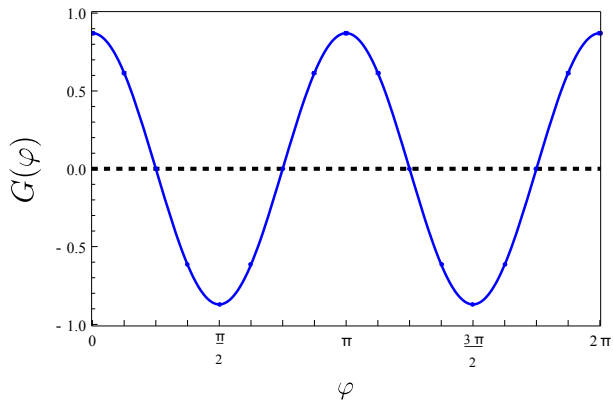}
  \end{center}
    \vspace{-0.3cm}
  \caption{{\bf Aspect's experiment and separability criterion}: The solid line represents $G(\varphi)$ defined in (7) measured by Aspect's experimental setup in 1981~\cite{aspect1}. The dashed line represents the prediction of a separable pure state in (2). } \label{fig01}
\end{figure}
The experiments, which we use to test saparability, have been performed in the past by Freedman and Clauser in 1972~\cite{freedman} and Aspect, Grangier and Roger in 1981~\cite{aspect1} (before the better known experiment in 1982~\cite{aspect2}). Those experiments are based on the measurement of the transverse linear polarization of the photon. To rewrite our relations, which are often written for spin operators, for the analysis of the photon measurement, one may define the projection operator such as $P({\bf a})=(1+{\bf a}\cdot {\bf \sigma})/2$ or ${\bf a}\cdot {\bf \sigma}=2P({\bf a})-1$. The projector $P({\bf a})$ in the transverse direction is then {\em formally} identified with the photon linear polarizer in the direction ${\bf a}$; CHSH inequality based on hidden-variables models is correctly described by this formal replacement, but the properties under the rotation group, for example, cannot be correctly described and a more careful treatment is required, as is explained later. We then obtain the quantity 
\begin{eqnarray}
G({\bf a},{\bf b})
&\equiv& 
\langle{\bf a}\cdot {\bf \sigma}\otimes {\bf b}\cdot {\bf \sigma}\rangle-\langle{\bf a}\cdot {\bf \sigma}\otimes {\bf 1}\rangle\langle{\bf 1}\otimes {\bf b}\cdot {\bf \sigma}\rangle\nonumber\\
&=&4[\langle P({\bf a})\otimes P({\bf b})\rangle
-\langle P({\bf a})\otimes{\bf 1}\rangle
\langle{\bf 1}\otimes P({\bf b})\rangle], 
\end{eqnarray}
corresponding to (2).
In terms of measured quantities in~\cite{aspect1} , $G(\varphi)=G({\bf a},{\bf b})$ is written as  
\begin{eqnarray}
G(\varphi)&=&4[\frac{R(\varphi)}{R_0}-\frac{R_1R_2}{R_0^2}]\nonumber\\
&=&(0.971-0.029)(0.968-0.028)0.984 \cos2\varphi
\end{eqnarray}
where $\varphi$ stands for the angle between ${\bf a}$ and ${\bf b}$. The quantities $R(\varphi),\ R_1, \ R_2$ and $R_0$ are defined in eq.(2) of~\cite{aspect1}, and the numerical factors which appear in front of $\cos2\varphi$ are also given in~\cite{aspect1}. See also Refs.~\cite{chsh, freedman}. 

Quantum mechanically, we have the prediction
\begin{eqnarray}
G({\bf a},{\bf b})=\cos2\varphi
\end{eqnarray}
for the (scalar) state 
\begin{eqnarray}
\psi=(1/\sqrt{2})[|H\rangle_{1}|H\rangle_{2}+|V\rangle_{1}|V\rangle_{2}] 
\end{eqnarray}
of linearly polarized (horizontal $|H\rangle$ and vertical $|V\rangle$) photons, which we expect for the cascade $6^{1}S_{0}\rightarrow 4{}^{1}P_{1}\rightarrow 4{}^{1}S_{0}$ in calcium~\cite{aspect1}; this state corresponds to $F_{\mu\nu}F^{\mu\nu}$  in the 4-dimensional  notation with $F_{\mu\nu}$ standing for the Maxwell field strength tensor, and proportional to $\vec{A}(1)\cdot \vec{A}(2)$ in terms of the transverse vector potential. The mean life of
the intermediate state $4{}^{1}P_{1}$ is about $4.5\times 10^{-9}$ sec~\cite{kocher}, and one may {\em assume} no decoherence and thus a pure or close to pure two-photon state in the present analysis. For the ideal measurement, the coefficient of $\cos2\varphi$ in (7) is expected to be close to unity. In fact, the authors in~\cite{aspect1} mentioned the good agreement of measured results with quantum mechanical predictions. 

We show the measured result of (7), which agrees well with the quantum mechanical prediction~\cite{aspect1}, in Fig.1 together with the prediction of separable state (2), namely, a separable pure quantum state. Fig.1 shows that the criterion (2) applied to (7) is very effective and provides a good evidence of inseparability of measured data by assuming a pure state. 
\\

\noindent {\bf Separability criteria of mixed states}:\\

One may still argue that there is no guarantee that the two-photon state in~\cite{aspect1} is a pure state which we assumed in the analysis of our Fig.1, although the good agreement of the observed data with the quantum mechanical prediction, as is noted in~\cite{aspect1}, and the very fact that one can make a quantum mechanical prediction indicates that the state is close to a pure state. To cope with this criticism and also to deal with general mixed states, we next
formulate convenient criteria to test the separability of general mixed states in (1). 

This formulation is motivated by the measurement of two-proton correlation by Sakai et al.~\cite{sakai}. They measure a pair of massive spin $1/2$ protons from the decay of the short-lived ($<10^{-21}$ sec) ${}^{2}$He spin-singlet state~\cite{sakai}. The initial spin  state is most likely a pure state $|\psi_{s}\rangle=(1/\sqrt{2})[|+\rangle_{1}|-\rangle_{2}-|-\rangle_{1}|+\rangle_{2}]$.  For this maximally entangled state, quantum mechanics predicts 
\begin{eqnarray}
G({\bf a},{\bf b})=-\cos\varphi, 
\end{eqnarray}
for the quantity in (6).
But the authors in~\cite{sakai} actually measured the quantity corresponding to the quantum mechanical correlation
\begin{eqnarray}
C_{QM}(\varphi)=\langle \psi|{\bf n}^{(1)}\cdot {\bf \sigma}\otimes {\bf n}^{(2)}\cdot {\bf \sigma}|\psi\rangle,
\end{eqnarray}
with unit vectors ${\bf n}^{(1)}$ and ${\bf n}^{(2)}$. 
They did not measure $\langle \psi|{\bf n}^{(1)}\cdot {\bf \sigma}\otimes {\bf 1}|\psi\rangle$ and $\langle \psi|{\bf 1}\otimes {\bf n}^{(2)}\cdot {\bf \sigma}|\psi\rangle$ separately, and thus we cannot construct the quantity $G({\bf n}^{(1)},{\bf n}^{(2)})$ defined in (6). They however mentioned that "Thus, a measured value of $C_{exp}(\varphi)=-\cos\varphi$, which is same as
$C_{QM}(\varphi)$, for ${\bf n}^{(1)}$ and ${\bf n}^{(2)}$ randomly rotated, is strong evidence that the incident two protons are in the entangled
state."  We thus define a quantity to test the general separable mixed state in (1) by incorporating the angular average in their measurement. 

We start with an arbitrary separable pure quantum state $|\psi\rangle=|{\bf s}_{1}\rangle|{\bf s}_{2}\rangle$ which gives $C_{QM}(\varphi)=({\bf s}_{1}\cdot {\bf n}^{(1)})({\bf n}^{(2)}\cdot {\bf s}_{2})$.  Here ${\bf s}_{1}$ and ${\bf s}_{2}$ stand for unit spin vectors which define the pure state by $|{\bf s}_{1}\rangle\langle{\bf s}_{1}|=\frac{1}{2}[1+{\bf s}_{1}\cdot{\bf \sigma}]$, for example. 
The prediction of the separable mixed state in (1) is thus written as 
\begin{eqnarray}
{\rm Tr}\rho\{ {\bf n}^{(1)}\cdot {\bf \sigma}\otimes {\bf n}^{(2)}\cdot {\bf \sigma}\}=\int d\Omega_{s_{1}}d\Omega_{s_{2}}w({\bf s}_{1},{\bf s}_{2})({\bf s}_{1}\cdot {\bf n}^{(1)})({\bf n}^{(2)}\cdot {\bf s}_{2})
\end{eqnarray}
where we use the continuum notation with $\int d\Omega_{s_{1}}d\Omega_{s_{2}}w({\bf s}_{1},{\bf s}_{2})=1$ instead of the discrete one, $\sum_{k}w_{k}=1$, in (1). This defines the most general separable mixed states and thus naturally incorporates the positivity of partial transposed density matrix. We emphasize that the geometrical average of the unit vector ${\bf n}$ covers all the possible spin states in the $d=2$ Hilbert space, which becomes crucial when one compares the spin average with the average over photon polarization later. This quantity becomes 
\begin{eqnarray}
C_{QM}(\varphi)_{mixed}&=&{\rm Tr}\rho \{{\bf n}^{(1)}\cdot {\bf \sigma}\otimes {\bf n}^{(2)}\cdot {\bf \sigma}\}_{ave}\nonumber\\
&=&\frac{1}{3}\int d\Omega_{s_{1}}d\Omega_{s_{2}}w({\bf s}_{1},{\bf s}_{2})({\bf s}_{1}\cdot {\bf s}_{2})\cos\varphi\nonumber\\
&=&\frac{1}{3}C\cos\varphi 
\end{eqnarray}
after the angular averaging of ${\bf n}^{(1)}$ and ${\bf n}^{(2)}$ with fixed $\cos\varphi={\bf n}^{(1)}\cdot{\bf n}^{(2)}$; note that the angular average means the average with solid angle, namely, $\int \cos^{2}\theta d\cos\theta d\phi/4\pi=1/3$.  It is interesting that the separable state gives a non-trivial $\cos\varphi$ dependence in $C_{QM}(\varphi)$. The quantity $C$ in (13) is defined by 
\begin{eqnarray}
C\equiv\int d\Omega_{s_{1}}d\Omega_{s_{2}}w({\bf s}_{1},{\bf s}_{2})({\bf s}_{1}\cdot {\bf s}_{2})
\end{eqnarray}
which satisfies $-1\leq C\leq 1$.
The relation (13) is the prediction of separable quantum states in (1).  The measured value in~\cite{sakai}, $C_{exp}(\varphi)=-\cos\varphi$ which is same as quantum mechanical prediction~\cite{sakai} (expected in their experimental setting), contradicts the prediction $C_{QM}(\varphi)_{mixed}$ and thus clearly shows inseparability. See Fig.2. 
\begin{figure}[h]
 \begin{center}
    \includegraphics[width=80mm]{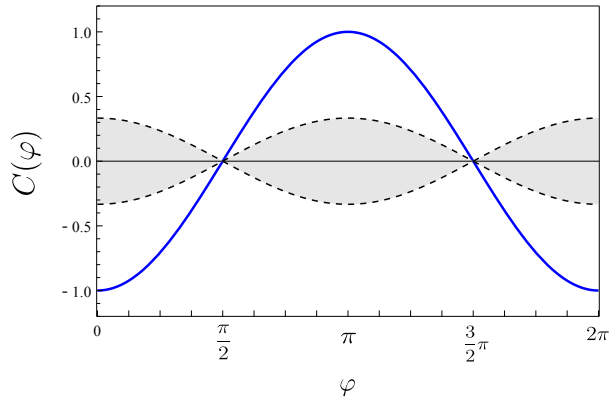}
 \end{center}
    \vspace{-0.3cm}
  \caption{{\bf Separable mixed state and Sakai's experiment}: Shaded region is the prediction of the separable mixed quantum state in eq.(13), and the solid line is the experimental result $C_{exp}(\varphi)=-\cos\varphi$ of Sakai et al.~\cite{sakai} }\label{fig02}
\end{figure}

Similarly,  one may consider the angular averaging of the photon case, but the {\em angular average} of the spinor and the photon is different. To analyze this issue,
we start with a general quantum state in the $d=2$ Fock space,
\begin{eqnarray}
|\alpha,\gamma\rangle=\cos\alpha e^{-i\gamma}\hat{a}^{\dagger}_{+}|0\rangle +\sin\alpha \hat{a}^{\dagger}_{-}|0\rangle 
\end{eqnarray}
which is parameterized by two real numbers $\alpha$ and $\gamma$. We have $[\hat{a}_{\pm},\hat{a}^{\dagger}_{\pm}]=1$ and $\hat{a}_{\pm}|0\rangle=0$. The second quantized operator for the photon propagating in $z$-direction with 
${\bf p}=(0,0,|{\bf p}|)$ is given by (in $d=2$ subspace by ignoring all the space-time indices)
\begin{eqnarray}
\hat{A}({\bf p})=u_{+}\hat{a}_{+}+u_{-}\hat{a}_{-}+ h.c.,
\end{eqnarray}
where one may choose the polarization vectors in the $x-y$ plane as 
\begin{eqnarray}
u_{+}=\left(\begin{array}{c}
            1\\
            0
            \end{array}\right),\ \ \ 
u_{-}=\left(\begin{array}{c}
            0\\
            1
            \end{array}\right),           
\end{eqnarray} 
without writing the vanishing $z$ component; $\pm$ states may be termed as $x$ and $y$ states or $H$ and $V$ states.
Note that ${\bf p}\cdot \hat{A}({\bf p})=0$ due to the Coulomb condition.
The wave function in the sense of the first quantization is given by 
\begin{eqnarray}
\psi(\alpha,\gamma)=\langle 0|\hat{A}({\bf p})|\alpha,\gamma\rangle=\cos\alpha e^{-i\gamma}u_{+}+\sin\alpha u_{-}.
\end{eqnarray}
We define the projection operator specifying the {\em measured linear polarization}  of 
the photon, which is chosen by the experimental setup, by 
\begin{eqnarray}
P(\theta)=|\theta,0\rangle\langle\theta,0|
\end{eqnarray}
using the state in (15) with $0\leq \theta\leq 2\pi$; namely $\alpha=\theta$ and $\gamma=0$ in (15). The general separable state may be defined in terms of the states in (15) by 
\begin{eqnarray}
\rho=\int d\Omega_{1}d\Omega_{2}w(\theta_{1}/2,\phi_{1};\theta_{2}/2,\phi_{2})
|\theta_{1}/2,\phi_{1}\rangle|\theta_{2}/2,\phi_{2}\rangle\langle\theta_{1}/2,\phi_{1}|\langle\theta_{2}/2,\phi_{2}|
\end{eqnarray}
using the "rotation" invariant volume element, which is analogous to the separable mixed spin states in (12), with $\int d\Omega_{1}d\Omega_{2}w(\theta_{1}/2,\phi_{1};\theta_{2}/2,\phi_{2})=1$. 
We now evaluate 
\begin{eqnarray}
&&4\langle\theta_{1}/2,\phi_{1}|\langle\theta_{2}/2,\phi_{2}|P(\theta_{a})\otimes P(\theta_{b})|\theta_{1}/2,\phi_{1}\rangle|\theta_{2}/2,\phi_{2}\rangle\\
&&=[1+\cos2\theta_{a}\cos\theta_{1}+\sin2\theta_{a}\sin\theta_{1}\cos\phi_{1}][1+\cos2\theta_{b}\cos\theta_{2}+\sin2\theta_{b}\sin\theta_{2}\cos\phi_{2}]\nonumber
\end{eqnarray}
which becomes 
\begin{eqnarray}
&&1+\frac{1}{2}[\cos\theta_{1}\cos\theta_{2}+\sin\theta_{1}\cos\phi_{1}\sin\theta_{2}\cos\phi_{2}]\cos2\varphi\nonumber\\
&&-\frac{1}{2}[\cos\theta_{1}\sin\theta_{2}\cos\phi_{2}-\cos\theta_{2}\sin\theta_{1}\cos\phi_{1}]\sin2\varphi
\end{eqnarray}
after angular averaging over $\theta_{a}$ and $\theta_{b}$ with fixed $\varphi
=\theta_{a}-\theta_{b}$.

We thus obtain 
\begin{eqnarray}
4{\rm Tr}\rho [P(\theta_{a})\otimes P(\theta_{b})]=1+\frac{1}{2}C\cos2\varphi
\end{eqnarray}
with
\begin{eqnarray}
C=\int d\Omega_{1}d\Omega_{2}w(\theta_{1}/2,\phi_{1};\theta_{2}/2,\phi_{2})[\cos\theta_{1}\cos\theta_{2}+\sin\theta_{1}\cos\phi_{1}\sin\theta_{2}\cos\phi_{2}]
\end{eqnarray}
which is bounded by $-1\leq C\leq 1$, if one recalls that ${\bf v}_{1}\cdot{\bf v}_{2}=
\cos\theta_{1}\cos\theta_{2}+\sin\theta_{1}\cos\phi_{1}\sin\theta_{2}\cos\phi_{2}$
when one defines a vector ${\bf v}=(\cos\theta, \sin\theta\cos\phi)$ whose magnitude is smaller than unity. The term with $\sin2\varphi$ in (22) vanishes if one assumes  that the positive semi-definite weight factor $w(\theta_{1}/2,\phi_{1};\theta_{2}/2,\phi_{2})$ contains only the components symmetric under $1\leftrightarrow 2$, which is a natural assumption. 

This prediction (23) of the separable mixed state in (1) that is formulated without assuming a pure state contradicts the experimental result given in (7)
\begin{eqnarray}
C_{exp}(\varphi)=0.996+0.88\cos2\varphi,
\end{eqnarray}
which agrees well with the quantum mechanical prediction~\cite{aspect1},
and thus the data clearly show inseparability. See Fig.3.
As for the justification of angular averaging, the authors in~\cite{aspect1} mentioned that, "we never observed any deviation from rotational invariance".
In passing, we mention that the two-photon state expected in the decay of a pseudo-scalar state ( a spin $0$ state with negative parity) $\epsilon^{\mu\nu\alpha\beta} F_{\mu\nu}F_{\alpha\beta}$ in the 4-dimensional  notation and proportional to $\vec{k}\cdot(\vec{A}(1)\times\vec{A}(2))$ in terms of the transverse potential,
\begin{eqnarray}
\psi_{ps}=(1/\sqrt{2})[|H\rangle_{1}|V\rangle_{2}-|V\rangle_{1}|H\rangle_{2}] 
\end{eqnarray}
gives
\begin{eqnarray}
C_{ps}(\varphi)=1-\cos2\varphi
\end{eqnarray}
in contrast to $C_{s}(\varphi)=1+\cos2\varphi$ for the scalar state in (9). This prediction is also tested by the criterion of separable mixed states in Fig.3.

\vspace{0.3cm}
\begin{figure}[h]
 \begin{center}
    \includegraphics[width=80mm]{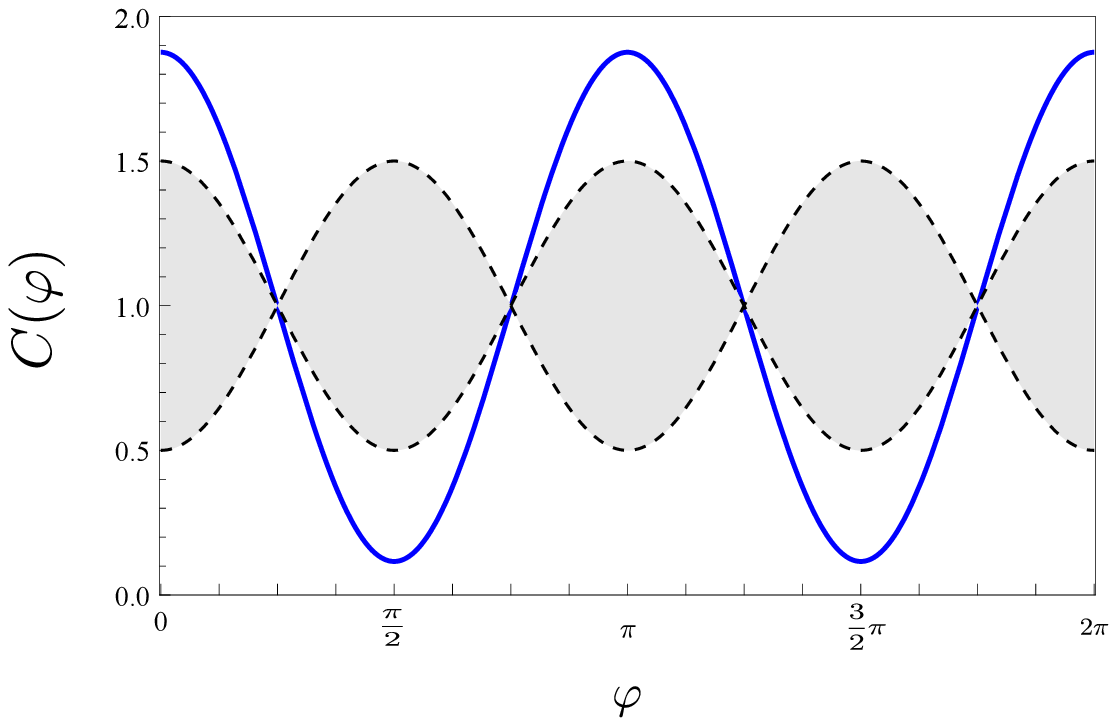}
 \end{center}
    \vspace{-0.3cm}
  \caption{{\bf Separable mixed state and Aspect's experiment in 1981}: Shaded region is the prediction of the separable mixed quantum state in eq.(23), and the solid line is the experimental result $C_{exp}(\varphi)=0.996+0.88\cos2\varphi$ of Aspect et al.~\cite{aspect1}. }\label{fig03}
\end{figure}

The difference in the numerical coefficients in front of $C$ in (13) and (23) arises from the fact that the {\em geometrical} angular averaging procedure we perform is 3-dimensional for the spin $1/2$ freedom while it is 2-dimensional for the linear polarization of the photon which is confined in the  plane perpendicular to the momentum direction. One may ask if it is possible to obtain $1/3$ for the photon? It is possible if one takes an average in the $d=2$ Hilbert space; one may define the projection operator for the linearly polarized photon using the states in (15) by 
\begin{eqnarray}
P(\theta/2,\phi)=|\theta/2,\phi\rangle\langle\theta/2,\phi|
\end{eqnarray}
instead of (19). One then obtains 
\begin{eqnarray}
&&4\langle\theta_{1}/2,\phi_{1}|\langle\theta_{2}/2,\phi_{2}|P(\theta_{a}/2,\phi_{a})\otimes P(\theta_{b}/2,\phi_{b})|\theta_{1}/2,\phi_{1}\rangle|\theta_{2}/2,\phi_{2}\rangle\nonumber\\
&&=[1+\cos\theta_{a}\cos\theta_{1}+\sin\theta_{a}\sin\theta_{1}\cos(\phi_{a}-\phi_{1})][1+\cos\theta_{b}\cos\theta_{2}+\sin\theta_{b}\sin\theta_{2}\cos(\phi_{a}-\phi_{2})]\nonumber\\
&&=[1+{\bf n}_{a}\cdot{\bf s}_{1}][1+{\bf n}_{b}\cdot{\bf s}_{2}]
\end{eqnarray}
if one defines two vectors ${\bf n}=(\cos\theta, \sin\theta\cos\phi,\sin\theta\sin\phi)$ and 
${\bf s}=(\cos\theta, \sin\theta\cos\phi,\sin\theta\sin\phi)$. If one takes the average over ${\bf n}_{a}$ and ${\bf n}_{b}$ with fixed 
\begin{eqnarray}
\cos\tilde{\varphi}\equiv{\bf n}_{a}\cdot{\bf n}_{b}
\end{eqnarray}
one obtains 
\begin{eqnarray}
&&1+\frac{1}{3}({\bf s}_{1}\cdot{\bf s}_{2})\cos\tilde{\varphi}
\end{eqnarray}
This is the same result as for the spin $1/2$ case in (13) and one obtains 
\begin{eqnarray}
4{\rm Tr}\rho [P(\theta_{a}/2,\phi_{a})\otimes P(\theta_{b}/2,\phi_{b})]=1+\frac{1}{3}C\cos\tilde{\varphi}
\end{eqnarray}
with
\begin{eqnarray}
C=\int d\Omega_{1}d\Omega_{2}w(\theta_{1}/2,\phi_{1};\theta_{2}/2,\phi_{2})({\bf s}_{1}\cdot{\bf s}_{2})
\end{eqnarray}
which is bounded by $-1\leq C\leq 1$. In the final expression in (32), one may replace 
$\cos\tilde{\varphi}\rightarrow \cos2\varphi$ to compare it with the previous expression (23). 

Physically, what is involved in the second procedure in (32) is to consider the average over the states
\begin{eqnarray}
|\theta/2,\phi\rangle=\cos\theta/2 e^{-i\phi}\hat{a}^{\dagger}_{+}|0\rangle +\sin\theta/2 \hat{a}^{\dagger}_{-}|0\rangle
\end{eqnarray}
by not only changing $\theta$ by geometric operations but also by applying the "phase shifter" to the state $\hat{a}^{\dagger}_{+}|0\rangle$ to generate the states with various $\phi$. Since we take an average over a larger class of states, the coefficient in (32) is made smaller to $1/3$. 
 
Those formulas (13) and (23) (and also (32)) provide convenient criteria to test separability without an explicit reconstruction of the state $\rho$ from the measured data. The angular averaging is required for these formulas to be a test of separability, but it does not appear to be difficult to implement the angular averaging in actual experiments, as the past experiments we discussed indicate~\cite{aspect1, sakai}.

We here mention the well-known Werner state~\cite{werner}, 
\begin{eqnarray} 
\rho_{w}=\frac{1}{8}{\bf 1} +\frac{1}{2}|\psi_{s}\rangle\langle\psi_{s}|
\end{eqnarray}
with the spin singlet state $|\psi_{s}\rangle=(1/\sqrt{2})[|+\rangle_{1}|-\rangle_{2}-|-\rangle_{1}|+\rangle_{2}]$, which accommodates a specific hidden-variables representation and  thus satisfies CHSH inequality. This state is rotation invariant and gives rise to 
\begin{eqnarray}
{\rm Tr}\rho_{w}({\bf a}\cdot {\bf \sigma}\otimes {\bf b}\cdot {\bf \sigma})=-\frac{1}{2}\cos\varphi 
\end{eqnarray}
which violates the separability criterion (13) and thus inseparable. The explicit example of the Werner state shows that local realism (local hidden-variables models) and quantum mechanical separability are logically independent notions.  

It is interesting to examine 
the Werner state when it is applied to two photon systems. If one chooses the state $|\psi_{s}\rangle$ in (35) as the photon state $|\psi_{ps}\rangle$ given in (26), one obtains
\begin{eqnarray}
4{\rm Tr}\rho_{w}(P(\theta_{a})\otimes P(\theta_{b}))=1 - \frac{1}{2}\cos2\varphi 
\end{eqnarray}
for which we cannot judge separability using the criterion (23). This failure of the separability criterion is analogous to the failure of CHSH inequality.  Instead, we need to consider the projector in (28) for the Werner state,
\begin{eqnarray}
4{\rm Tr}\rho_{w}(P(\theta_{a}/2,\phi_{a})\otimes P(\theta_{b}/2,\phi_{b}))=1 - \frac{1}{2}\cos\tilde{\varphi} 
\end{eqnarray}
and compare this expression with the separability criterion (32) to judge the inseparability of the Werner state. (If one uses the photon state $|\psi\rangle$ in (8), one obtains $1 + \frac{1}{2}\cos2\varphi $ in (37) and $1 + \frac{1}{2}\cos\tilde{\varphi} $ in (38), respectively.) This analysis suggests that  we need to consider a larger class of  detector states in (28) to detect inseparability of the Werner state by photon measurements,. 
    
\section{Discussion}
As for the test of separability of general mixed quantum states in (1), Peres criterion of the positivity of  partial transposed density matrix 
\begin{eqnarray}
\rho^{T_{2}}=\sum_{ijkl}P^{ij}_{kl}|i\rangle\langle j|\otimes|l\rangle\langle k|
\end{eqnarray}
for the original $\rho=\sum_{ijkl}P^{ij}_{kl}|i\rangle\langle j|\otimes|k\rangle\langle l|$, which gives a necessary and sufficient separability condition for $d=2\times 2$ systems~\cite{peres}, can be used in combination with the state reconstruction of two linearly polarized photons such as in~\cite{james}. But in practice the reconstruction of the state $\rho$ from the measured data is generally involved~\cite{james}. We have instead presented simpler formulas (13) and (23), which incorporate the angular averaging of measured two-point correlations while assuming the most general separable density matrix.
Our simplified test of separability nicely works without any extra cost such as the state reconstruction, and the final outcome in Fig.1 - Fig.3 may be favorably compared with the standard test of CHSH inequality~\cite{aspect1,sakai}, although our formula tests separability while CHSH inequality tests locality and thus not identical. From the point of view of separability, CHSH inequality and our relation (2) both give the necessary and sufficient condition for any pure state, while CHSH inequality and our relations (3) and (4) both give only the necessary condition for mixed states. The Werner state is tested by our criterion in the case of spin, but in the case of photon one needs the information coming from the Hilbert space averaging.

We here mention a past work on the test of entanglement~\cite{yu} which is closely related to the present study. A comprehensive and comparative study of various separability criteria has been given in~\cite{acin}. The authors in~\cite{yu}  show that the separability of the density matrix for a two-spin system is characterized by an algebraic inequality derived from Robertson's uncertainty relations for Pauli matrices. The algebraic inequality is shown to give a necessary and sufficient separability criterion for the two-spin system in $d=2\times2$, and thus the inequality can be used in place of the positivity of partial transposed density matrix. However, it remains hard to confirm the separability of an unknown state in practice because one has to check the proposed inequality for all sets of local complementary (non-commuting) observables, despite of the fact that the analysis in~\cite{yu} ensures in principle a violation of the inequality for any entangled state by choosing properly the local testing observables.  

The authors of~\cite{yu} also suggest a simple {\em necessary} condition of separability by just measuring 3 correlations $\langle\sigma^{1}\otimes\sigma^{1}\rangle, \ \ \langle\sigma^{2}\otimes\sigma^{2}\rangle, \ \ \langle\sigma^{3}\otimes\sigma^{3}\rangle$ and examining 
\begin{eqnarray}
-1\leq \langle\sigma^{1}\otimes\sigma^{1}\rangle +\langle\sigma^{2}\otimes\sigma^{2}\rangle +\langle\sigma^{3}\otimes\sigma^{3}\rangle\leq 1 
\end{eqnarray}
in the case of spin $1/2$ systems. This condition is based on the equality
\begin{eqnarray}
({\bf s}_{1}\cdot {\bf s}_{2})
=\langle{\bf s}_{1}|\langle{\bf s}_{2}|[\sigma^{1}\otimes\sigma^{1}+\sigma^{2}\otimes\sigma^{2}+\sigma^{3}\otimes\sigma^{3}]|{\bf s}_{1}\rangle|{\bf s}_{2}\rangle,
\end{eqnarray}
which is bounded by unity.
The expected initial spin  state $|\psi_{s}\rangle=(1/\sqrt{2})[|+\rangle_{1}|-\rangle_{2}-|-\rangle_{1}|+\rangle_{2}]$ in~\cite{sakai}
gives $\langle\psi_{s}|\sigma^{1}\otimes\sigma^{1}|\psi_{s}\rangle=\langle\psi_{s}|\sigma^{2}\otimes\sigma^{2}|\psi_{s}\rangle=\langle\psi_{s}|\sigma^{3}\otimes\sigma^{3}|\psi_{s}\rangle=-1$, and this comparison shows the inseparability of the state $|\psi_{s}\rangle$ and provides information to understand the criterion at $\varphi=0$ in Fig.2. This analysis, when applied to the test of the inseparability of the Werner state $\rho_{w}$ in (35), leads to the same conclusion as  our relation (36) at $\varphi=0$; it is known that $\beta\leq 1/3$  is the necessary and sufficient condition of separability if one replaces the coefficient $1/2$ of the singlet state appearing in the Werner state (35) by $\beta$~\cite{werner}, namely,
$\rho_{w}=\frac{1}{4}(1-\beta){\bf 1} +\beta|\psi_{s}\rangle\langle\psi_{s}|
$.

For the photon case instead of the spin $1/2$ case discussed in (41), one can show the relation 
\begin{eqnarray}
&&\psi^{\dagger}(\theta_{1}/2,\phi_{1})\psi^{\dagger}(\theta_{2}/2,\phi_{2}) [\sigma^{1}\otimes\sigma^{1}+\sigma^{3}\otimes\sigma^{3}]\psi(\theta_{1}/2,\phi_{1})\psi(\theta_{2}/2,\phi_{2})\nonumber\\
&&=\cos\theta_{1}\cos\theta_{2}+\sin\theta_{1}\cos\phi_{1}\sin\theta_{2}\cos\phi_{2},
\end{eqnarray}
in terms of separable wave functions in (20), which predicts 
\begin{eqnarray}
-1\leq \langle\sigma^{1}\otimes\sigma^{1}\rangle + \langle\sigma^{3}\otimes\sigma^{3}\rangle\leq 1 
\end{eqnarray}
for the general separable states in (20). Here Pauli spinors act on the qubit states defined by the photon. In comparison, one can confirm the relation $\psi^{\dagger} [\sigma^{1}\otimes\sigma^{1}+\sigma^{3}\otimes\sigma^{3}]\psi=2$ for the singlet (scalar) wave function $\psi$ in (8), and $\psi_{ps}^{\dagger} [\sigma^{1}\otimes\sigma^{1}+\sigma^{3}\otimes\sigma^{3}]\psi_{ps}=-2$ for the pseudo-scalar wave function $\psi_{ps}$ in (25). The comparison of these relations shows the inseparability of both $\psi$ in (8) and $\psi_{ps}$ in (25), and it provides information to understand the curves at the point $\varphi=0$ in Fig.3 (if one corrects the overall shift by unity), although the geometrical meaning of the correlations in (43) in the real 3-dimensional space is not obvious.

It may be interesting to note that the analysis in~\cite{yu} is closely related to the analysis by Simon~\cite{simon} who studied the separability criterion for the two-party system with a one-dimensional continuous freedom in each party by writing an algebraic inequality based on the positivity of partial transposed density matrix; the number of independent linear operators is 4 since the phase space dimensionality is $d=2\times2$ in~\cite{simon}.  The criterion, which provides a  necessary and sufficient separability condition for Gaussian states in quantum optics~\cite{simon, duan}, has been later shown to agree with an inequality based on Kennard's uncertainty relation~\cite{fujikawa}. It appears that the Heisenberg uncertainty principle accounts most of separability criteria in lower dimensional systems.

The test of separability and the test of local realism are closely related in that both test entanglement  but there is a crucial difference. The separability is based on quantum mechanics while the test of local realism is based on local hidden-variables models and thus generally tests the deviation from quantum mechanics also. The probability interpretation of quantum mechanics and hidden-variables models are based on the notion of {\em valuation}.
The valuation is defined as the probability measure $v$ which assigns non-negative values $v(P_{k})$ to any set of complete orthogonal projection operators $\sum_{k}P_{k}=1$
by preserving the linearity condition $v(\sum_{k}P_{k})=\sum_{k}v(P_{k})=1$. A general analysis~\cite{gleason, kochen, beltrametti} shows that such a measure is inevitably given by a trace representation $v(P_{k})=Tr (\rho P_{k})$ with a suitable trace-class operator $\rho$~\cite{neumann} for the dimensions of the Hilbert space $d\geq 3$, and no deterministic (dispersion-free $v(P^{2}_{k})-v(P_{k})^{2}=0$) representation of $v$ such as in hidden-variables models is possible for $d\geq 3$. 
There is also an analysis~\cite{busch} which asserts that, if one adopts POVMs, $\sum_{k}E_{k}=1$, instead of projection operators, those $\{E_{k}\}$ are
jointly measurable in a single experiment~\cite{busch}  but no deterministic representation of $v$ with $v(\sum_{k}E_{k})=\sum_{k}v(E_{k})=1$ (by maintaining the dispersion-free condition $v(E^{2}_{k})-v(E_{k})^{2}=0$) is possible even in $d=2$, while a trace representation $v(E_{k})=Tr (\rho E_{k})$ of von Neumann is always possible. If one should adopt this last point of view, no sensible local non-contextual hidden-variables models in $d=2\times2$ would be defined. In such a case, the separability provides an indispensable  test of entanglement.

\section*{Acknowledgments}
One of the authors (K.F.) thanks K-K. Phua for the hospitality at IAS, Nangyang Technological University. We also thank N. Namekata for calling a useful experimental reference to our attention.
This work is partially supported by JSPS KAKENHI (Grant No. 25400415) and the National Research Foundation and Ministry of Education, Singapore (Grant No. WBS: R-710-000-008-271).


\begin{thebibliography}{99}
\bibitem{bell}
J. S. Bell, Physics {\bf 1}, 195 (1964).
\bibitem{chsh}
J. F. Clauser, M. A. Horne, A. Shimony and R. A. Holt, Phys. Rev. Lett. 
{\bf 23},  880 (1969).
\bibitem{peres}
A. Peres, Phys. Rev. Lett. {\bf 77}, 1413 (1996).\\
P. Horodecki, Phys. Lett. A{\bf 232}, 333 (1997).
\bibitem{weigert}
St. Weigert, "Reconstruction of Spin States and its Conceptual Implications", in
{\em New Insights in Quantum Mechanics}, edited by H.-D. Doebner, S. T. Ali, M. Keyl, and R. F. Werner, (World Scientific, 1999). arXiv:quant-ph/9809065v1.
\bibitem{james}
D.F.V. James et al, Phys. Rev. A{\bf 64}, 052312 (2001).
\bibitem{aspect1}
A. Aspect, P. Grangier, G. Roger, Phys. Rev. Lett. {\bf 47}, 460 (1981).
\bibitem{sakai}
H. Sakai et al., Phys. Rev. Lett. {\bf 97}, 150405 (2006).
\bibitem{werner}
R.F. Werner, Phys. Rev. A{\bf 40}, 4277 (1989).
\bibitem{freedman}
S.J. Freedman and J.F. Clauser, Phys. Rev. Lett. {\bf 28}, 938 (1972).
\bibitem{aspect2}
A. Aspect, J. Dalibard and G. Roger, Phys. Rev. Lett. {\bf 49}, 1804
(1982).
\bibitem{kocher}
C. A. Kocher and E. D. Commins, Phys. Rev. Lett. {\bf 18}, 575
(1967).
\bibitem{yu}
Sixia Yu, Jian-Wei Pan, Zeng-Bing Chen, and Yong-De Zhang, Phys. Rev. Lett. {\bf 91},
217903 (2003).
\bibitem{acin}
J. B. Altepeter,et al., Phys. Rev. Lett. {\bf 95}, 033601 (2005).
\bibitem{simon}
R. Simon, Phys. Rev. Lett. {\bf 84}, 2726 (2000).
\bibitem{duan}
L.M. Duan, G. Giedke, J.I. Cirac and P. Zoller, Phys. Rev. Lett. {\bf 84}, 2722 (2000). 
\bibitem{fujikawa}
K. Fujikawa, Phys. Rev. A{\bf 80}, 012315 (2009).
\bibitem{gleason}
A. M. Gleason, J. Math. Mech. {\bf 6}, 885 (1957).
\bibitem{kochen}
S. Kochen and E. P. Specker, J. Math. Mech. {\bf 17}, 59 (1967).
\bibitem{beltrametti}
E. G. Beltrametti and G. Gassinelli, {\em The Logic of Quantum 
Mechanics}, (Addison-Wesley Pub., Reading, 1981).
\bibitem{neumann}
J. von Neumann, {\em Mathematical Foundations of Quantum Mechanics}
(Princeton Univ. Press, Princeton, 1955).
\bibitem{busch}
P. Busch,  Phys. Rev. Lett. {\bf 91}, 120403 (2003).
\end{thebibliography}
\end{document}